# Human Body-Sway Steady-State Responses to Small Amplitude Tilts and Translations of the Support Surface - Effects of Superposition of the Two Stimuli

Vittorio Lippi[1,2,3], Christoph Maurer[2], Thomas Mergner[2]

## Abstract
### Background

For humans, control of upright standing posture is a prerequisite for many physical activities. Experimentally, this control is often challenged by the motion of the support surface presented as tilt or translation, or some combination thereof. In particular, we have investigated subjects balancing in situations where tilt and translation stimuli were presented in isolation and compared it to a situation where such stimuli occurred simultaneously.

### Research questions

Is the human posture control system in the case of two or more superimposed external disturbances responding to these as if it were dealing with one disturbance? Or does it identify the disturbances individually and as such and respond to them specifically, as suggested in a current concept of disturbance-specific estimations and compensations?

### Methods

We had healthy human subjects controlling their balancing of upright stance on a motion platform while we presented them with different combinations of pseudorandom support surface tilt and translation stimuli alone or in superposition (with peak-to-peak amplitude of 0.5° and 1° for tilt, and 0.8 cm and 1.5 cm for translation). In one set of trials they kept their eyes closed and in a second set open. Furthermore, a simulation was performed to qualitatively evaluate the impact of sensory non-linearities and joint stiffness modulation.

### Results

We found that the experimental conditions 'eyes open' vs. 'eyes closed' always created significant differences ($p<0.05$) between the frequency response functions. In contrast to this, with different combinations of the tilt and translation stimuli, significant differences between the responses were observed only in 5 cases over the 24 that have been tested.

### Significance

The superposition of translation and tilt can be used to characterize the responses to both stimuli with one trial. When the amplitude of the stimuli is unbalanced (e.g. very small tilt superimposed with a larger translation) the effect of stiffness modulation can be studied.

---

[1] Institute of Digitalization in Medicine, Faculty of Medicine and Medical Center - University of Freiburg, Freiburg, Germany

[2] Clinic of Neurology and Neurophysiology, Medical Centre-University of Freiburg, Faculty of Medicine, University of Freiburg, Germany

[3]  Corresponding author.

Breisacher Str. 64 79106 Freiburg im Breisgau (BW) Germany

ORCID Id: 0000-0001-5520-8974

vittorio.lippi@uniklinik-freiburg.de

# 1. Introduction

Human control of standing posture may experimentally be challenged by support surface motion, often presented as support surface tilt, translation, or some superposition of both. This study investigates how the human balance control mechanisms deal with such disturbances in the body's sagittal plane when they are presented in isolation versus in a superimposed way. The particular aim was to investigate how and to what extent the responses evoked by the two disturbances influence each other.

Support surface tilts as well as translations have extensively been used before in posture control research since they allow us to investigate and characterize the involved sensorimotor control channels in a measurable and repeatable way using input/output functions. Human posture and balance control rely mainly on combining vestibular, proprioceptive, and visual cues (Horak & Macpherson, 1996). Depending on the experimental scenarios, the use of these inputs is thought to get modulated in terms of *sensory reweighting* (Cenciarini, 2006; Nashner & Berthoz, 1978) i.e. changing the relative weight associated with the different sensory channels in the feedback control of upright stance. Specifically, *sensory reweighting* is observed with different sensory input availability depending on applied conditions (e.g. open vs closed eyes as in Assländer & Peterka, 2016) or medical conditions (e.g. Caccese et al., 2021, or Bonan et al., 2013), or different stimulus modalities (Cenciarini, 2006). Notably, however, in the presence of both, tilt and translation, the same sensory input may have a different functional significance. For example, without being integrated with vestibular input, the proprioceptive inputs from the ankle joints alone do not allow a subject to distinguish whether a given change in body orientation in space owes to a support surface tilt or translation. Experiments comparing vestibular loss subjects with vestibular able subjects indicated a differentiating role of the vestibular input. With support surface tilt as the stimulus, for example, the two subject groups responded differently (Thomas Mergner et al., 2009), whereas there was no significant difference in the responses of such groups with support surface translation (Lippi et al., 2020). Furthermore, the two disturbance modalities require different and, to a certain degree, even contradictory compensation strategies: With support surface translation, the body tends to become destabilized by its inertia, with joint stiffness helping to balance, whereas body inertia supports body stabilization with fast support surface tilt. On this background, we studied here the responses to tilt and translation disturbances both separately as well as in superposition. In particular, the present study tests experimentally to what extent the response to one of the two disturbance modalities, tilt and/or translation, exerts an effect on the response to the respective other disturbance modality, and accordingly discusses possibly observed differences between the two conditions. Previous studies (Allum et al., 1993, 1994; Allum & Honegger, 1998) tested the superposition of tilt and translation with short stimuli (200 ms or less) to study the early muscle response. In the present work, the steady-state response is analyzed using a continuous pseudorandom ternary sequences (PRTS) profile for translation stimuli allowed us to describe the body sway responses in terms of *frequency response functions*, i.e. as empirical transfer functions (Peterka, 2002). The disturbance peak-to-peak amplitudes proposed in this work are relatively small: tilt up to 1°, and translation up to 1.8 cm, in contrast to previous studies where larger disturbances were applied, for example, 8° tilt in (Georg Hettich et al., 2014) and 24 cm translation in (Joseph Jilk et al., 2014). The choice of small-amplitude stimuli is motivated by safety reasons and because it allows for comparison with previous studies that involved elderly or balance-impaired subjects (Davidson et al., 2011; Isabella Katharina Wiesmeier et al., 2015), thus with subjects who represent a relevant demographic for this kind of experiments (and the potential clinical use of such tests). Considering that the availability of vision has a significant effect on postural responses to support surface movements (Akçay et al., 2021; Assländer et al., 2015; Jilk et al., 2014), the interaction between the two stimulus modalities might have been affected as well. The stimulus scenarios are hence tested with both the eyes-open (EO) and the eyes-closed (EC) condition. This allowed us to evaluate the effect of vision from the difference between the responses with EO and EC, and simultaneously to evaluate how the responses to superimposed tilt and translation stimuli are affected.

A simulation of the EC case is then used to qualitatively address the hypothesis that different experimental scenarios, i.e. different combinations of the two disturbance modalities, are associated with different degrees of ankle joint stiffness, and to discuss the effect of nonlinearities in the human sensor fusions. As the results showed that the interaction between tilt and translation is similar in the two cases, the EO case was not simulated to keep the model simpler and the exposition shorter. The notion that support surface tilt produces smaller ankle stiffness than support surface translation is suggested by the role of the ankle in transmitting the tilt perturbation to the body: In the ideal case with tilt and no ankle stiffness, the body would tend to maintain by its inertia the upright position. From an experimental point of view, a reduction in joint stiffness as an adaptation to support surface tilt has previously been hypothesized after observing the behavior of healthy subjects before and after training in upright standing with such an external stimulus (Assländer et al., 2020). A related, yet opposite hypothesis can be forwarded for support surface translations, where body inertia destabilizes the equilibrium, with the ankle stiffness then helping the balancing. Besides the joint stiffness associated with mechanical properties of the ankle, also short-latency monosynaptic stretch reflex responses can contribute to balance in presence of support surface translations and destabilize the body during initial perturbation. Stretch reflexes can be modulated in response to predictable perturbation (L. M. Nashner, 1976). It is possible that, although the perturbation used here has a pseudorandom profile and hence it is not directly predictable, the subject can modulate reflexes according to the scenario as it is observed that an improvement of the responses with learning can happen similarly with rhythmic and pseudorandom sequences (Assländer et al., 2020). This modulation may conflict when the translations and rotations are combined. In this work, the passive joint stiffness accounts for phenomena such as co-contractions and reflexes with a delay that is negligible compared with the one associated with the feedback loop, a hypothesis about the modulation of such passive stiffness will be formulated in § 2.5, *Modelling and Simulation* to be compared with the observed responses.

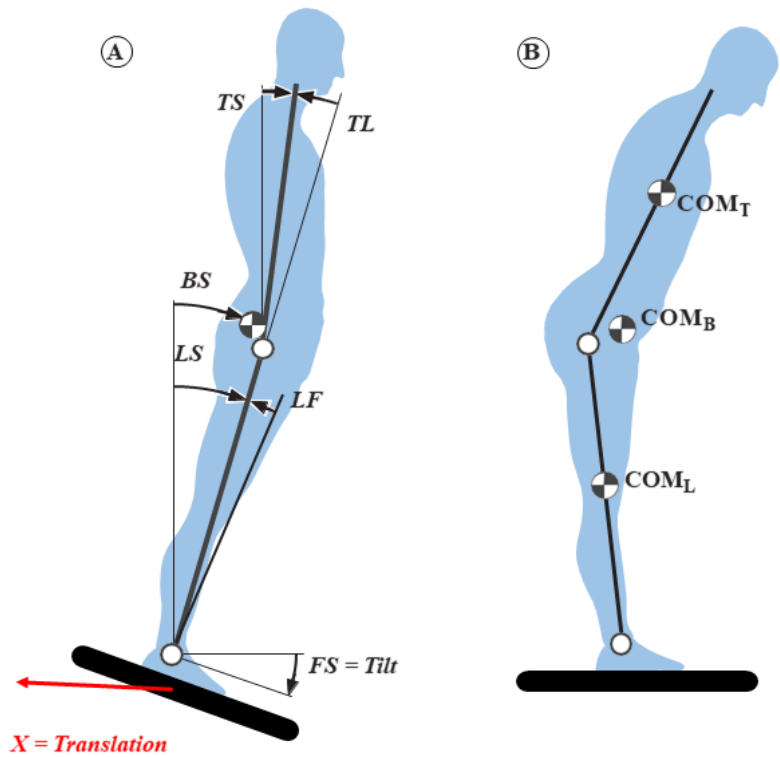

*Fig. 1(A) Human body kinematics and conventions for biomechanical variables describing the double inverted pendulum (DIP) and single inverted pendulum (SIP) models for balancing in the sagittal plane. The trunk segment includes also the head and the arms (not shown). The lower segment is given by the legs without the feet (feet are moving together with the support surface without loss of contact). Angles are expressed with respect to the gravitational vertical: the trunk-space angle TS and the leg-space angle LS. The foot-space angle FS equals the support surface tilt (FS= Tilt) and X represents the translation. The hip joint angle is defined by the trunk–leg angle TL and the ankle angle, here as leg–foot angle LF. (B) Position of the whole body center of mass ($COM_B$). $COM_T$ and $COM_L$ refer to the COM of the trunk segment and the leg segment, respectively. The SIP model used in the analysis considers the dynamics of BS (the sway of $COM_B$ with respect to the ankle joint). BS is computed on the basis of TL and LS that are tracked by an optical system, as explained in the Methods section.*

## 2. Materials and Methods
### 2.1. Experimental setup

A group of 32 healthy subjects (16 females, 16 males, aged between 22 and 58 years, average 36.75±11.28) was tested. The subjects were presented with stimuli consisting of motions of the support surface on which they were standing: Translation, tilt, and superposition of translation and tilt in the body-sagittal plane. The response tracking was performed using active markers (Optotrak 3020; Waterloo, ON, Canada), attached to subjects' hips and shoulders and to the platform, which allowed for the reconstruction of the body segment sways (TL, LS, and FS in Fig. 1). A custom-made program was used to generate the support surface motion. The marker positions were recorded at 100 Hz using software written in LabView (National Instruments; Austin, TX, United States). The profile used for the stimuli was a pseudorandom ternary signal, PRTS that was proposed by Peterka (2002), where it was used with different tilt amplitudes, but not for translation. The peak-to-peak amplitudes of the stimuli were set to 1° and 0.5° for the tilt and 1.52 cm and 0.76 cm for the translation (referred to in the following text by approximation, as 1.5 and 0.8 cm respectively). The intensity of these stimuli is small compared to what a healthy human subject can cope with. This approach was motivated by our aim to provide a data-set that we can compare with previously obtained sets from both healthy subjects *and* patients, e.g. in Wiesmeier et al. (2017) and Wiesmeier et al. (2015).

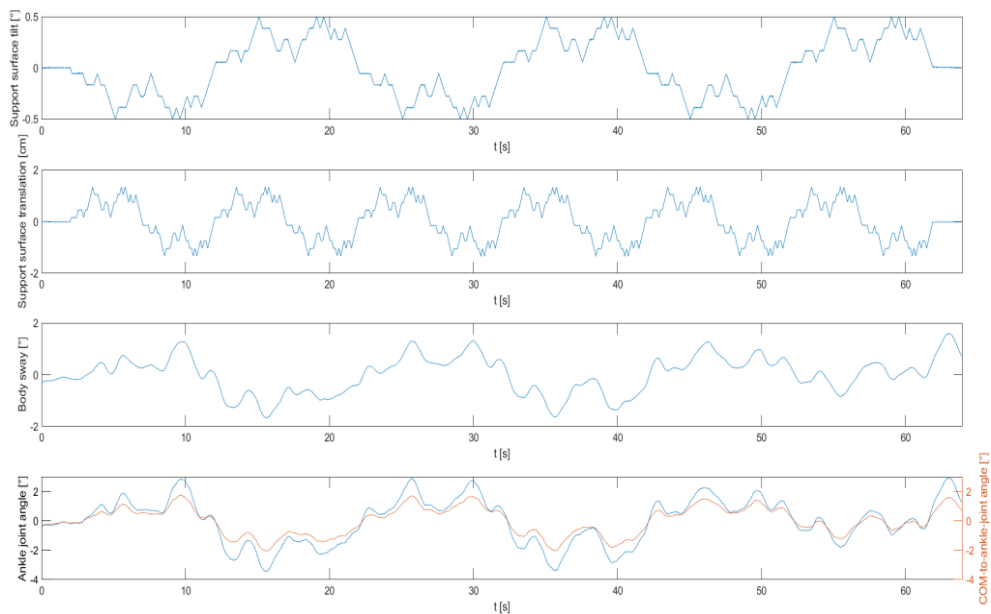

*Fig. 2. Example of profiles of support surface tilt (top) and translation (middle) stimuli, body-sway responses in space (third row), and ankle joint angle, leg-to-foot, in comparison with body-to-foot angle (bottom) from one set of the trials. The body sway is produced by the superimposition of tilt and translation shown above. With the small stimuli provided the profile of the COM sway with respect to the foot resembles the one of the ankle joint angle, with a slightly smaller amplitude*

## 2.2. Body tracking and kinematics

The body kinematics considered in the present study is shown schematically in Fig. 1 together with the conventions we used to refer to the body angles. The responses to the presented stimuli are described in terms of body-sway in space (BS), with the rotation angle of the body center of mass ($COM_B$) around the ankle joints expressed with respect to the gravitational vertical. To reconstruct the COM sway, the same set-up as in previous experiments (Assländer et al., 2015; G Hettich et al., 2011) was used: The optical tracking provides the hip and shoulders displacements, from which the sway of the upper body about the hip joint (BS) and of the lower body about the ankle joint (LS) was computed. Specifically, the lower body in space angle (LS) was calculated from the horizontal hip displacements and the height of the hip marker (measured manually before the beginning of the trials). The upper body angle in space (BS) was calculated from the height difference between hip and shoulder markers during the upright standing position of a subject and the difference in horizontal displacements of the two markers. An additional marker was fixed on the moving platform. The obtained marker inputs allowed for the reconstruction of the $COM_B$ position on the basis of human anthropometrics (see Winter, 2009). The COM sway is a good representation of the subjects' responses to small stimuli. Subjects exhibited an *ankle-strategy* behavior that can be modeled as a single inverted pendulum balancing thanks to ankle torque. In this scenario, the sway of the body segments is strongly correlated as shown by the example in Fig. 2 (bottom). Specifically, the correlation coefficient between body sway and ankle joint angle computed over the whole dataset is $r = 0.9894$.

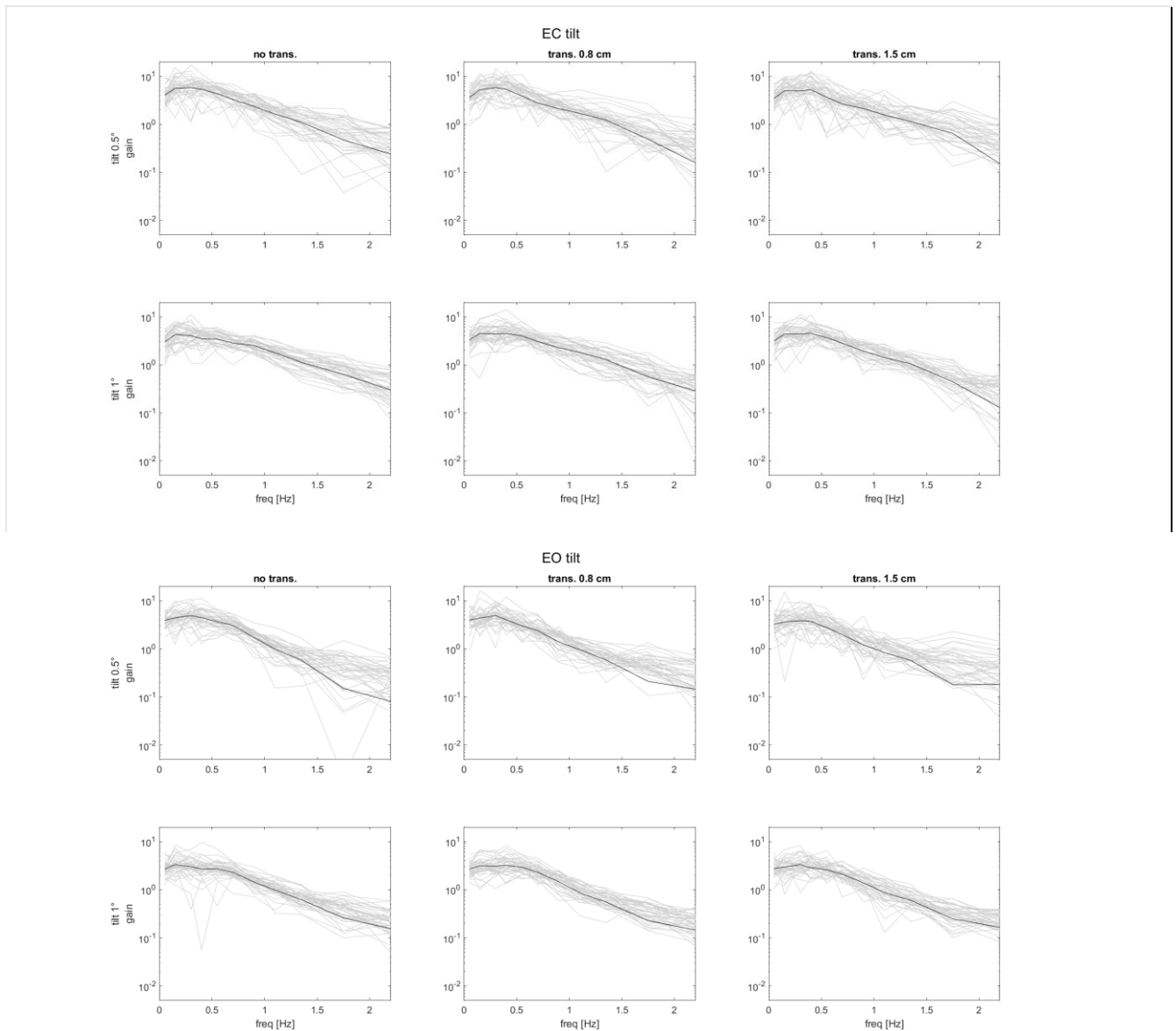

*Fig. 3 Support surface tilt to body sway FRF magnitudes from single trials (gray) and average responses (black). The panels on top show results from trials obtained with eyes closed (EC) and the ones at the bottom with eyes open (EO). The responses in the rows correspond to trials with the same stimulus amplitude (0.5° or 1°), while the columns cover the different amplitudes of the support surface translation: absent ('no trans'), 0.8 cm, and 1.5 cm.*

### 2.3. Frequency Response Function

The frequency response function, FRF, is an empirically computed transfer function between the stimulus (input) and the body sway (output). In detail, sway responses are averaged across all PRTS sequence repetitions across subjects, discarding the first cycle of each trial to avoid transient response effects. Spectra of the corresponding stimuli and body sway responses in space are computed using Fourier transforms. Finally, frequency response functions are computed as cross-power spectra $G_{xy}(f)$ divided by the stimulus power spectra $G_{yy}(f)$:

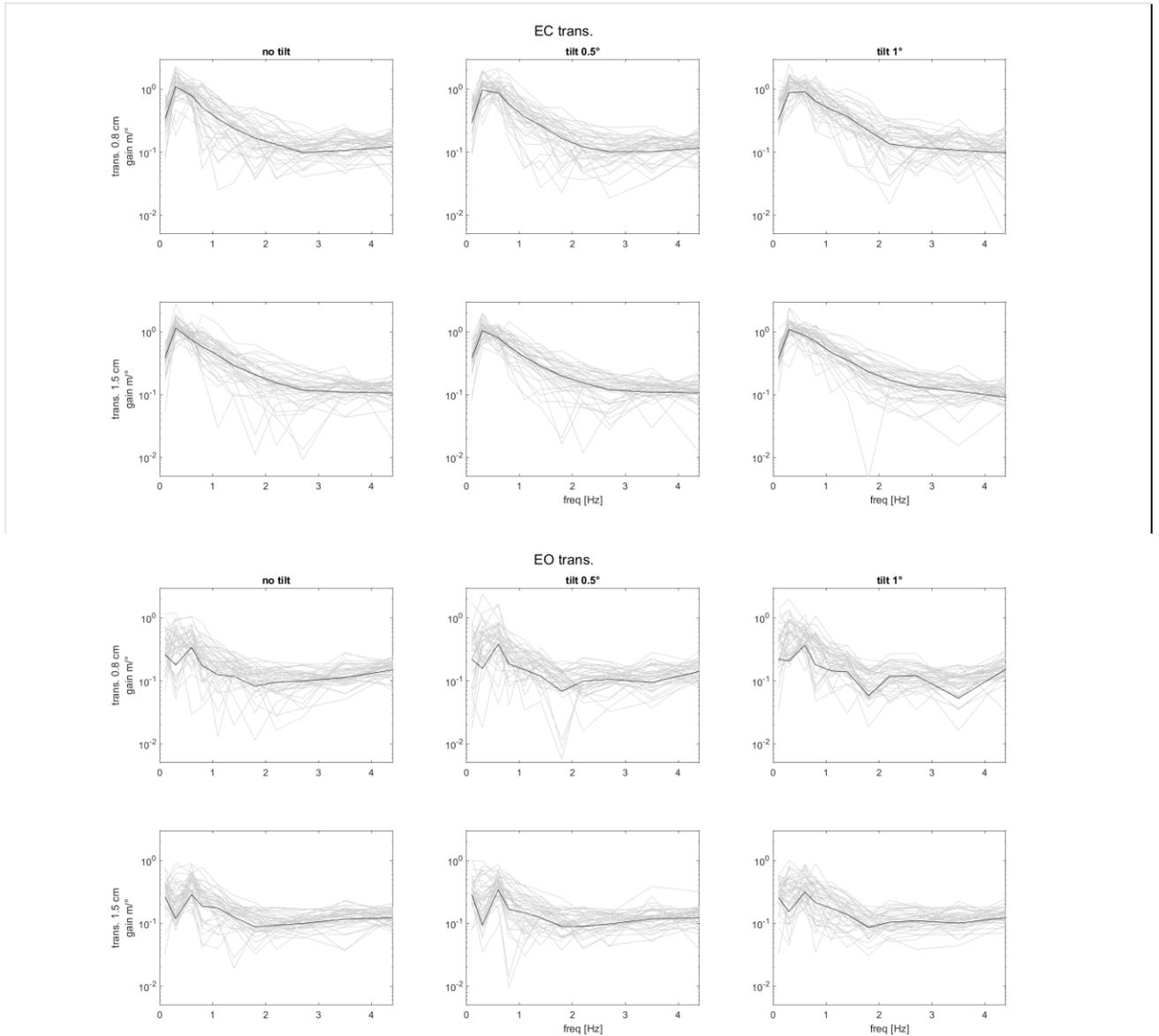

*Fig. 4.* Support surface translation to body sway FRF magnitude from single trials (gray) and average response (black). The figure on top shows trials obtained with the eyes closed condition (EC) and the one on the bottom the trials obtained with the eyes open (EO). The responses in each row correspond to trials with the same translation stimulus amplitude (0.8 cm or 1.5 cm) while those in the columns give the respective amplitudes of the support surface tilt: absent ('no tilt'), 0.5°, and 1°.

$$\frac{G_{xy}}{G_{yy}}$$

The values of the obtained transfer functions are averaged over bands of frequencies as shown before in Lippi et al. (2021), with the resulting FRFs being represented by a vector of 11 complex values.

In the present experiment, the profiles of tilt and translation stimuli were designed in a way so that there is no overlapping of the peaks of the power-spectrum in a given profile with those of the others since they fall on the zero regions of the power-spectra of the latter (Lippi et al., 2020; Peterka, 2002). This is achieved using a support translation profile similar to the tilt profile, but performed at twice the speed and repeated twice, as shown in Fig. 2. As a result, the FRFs for tilt and translation are plotted with respect to the following frequency points

$f_{tilt} = [0.05 \quad 0.15 \quad 0.3 \quad 0.4 \quad 0.55 \quad 0.7 \quad 0.9 \quad 1.1 \quad 1.35 \quad 1.75 \quad 2.2]$,

and $f_{trans} = 2f_{tilt} = [0.1\ \ 0.3\ \ 0.6\ \ 0.8\ \ 1.1\ \ 1.4\ \ 1.8\ \ 2.2\ \ 2.7\ \ 3.5\ \ 2.2]$, respectively.

The property of having non-overlapping spectra may not be immediately obvious when looking at the 11 frequencies, because some frequencies (i.e. 1.1 Hz and 2.2 Hz) are present for both the FRFs. The overlap is only apparent though because the frequency points used to compute the averages are in each case not coincident. The FRFs from the trials analyzed in this study are shown in Fig. 3 and Fig. 4 for the response to tilt and translation, respectively. Each FRF refers to a given trial, whereas a trial produces two FRFs if two stimuli are used (one for tilt and the other for translation). As the response to a combination of support surface tilt and translation is the sum of the FRFs multiplied by the respective inputs the time, and the FRFs are never negligible in the present experiment, the time domain body sway response is always affected by the presence of one of the two modalities. This is similar to what was observed in previous works with short-duration stimuli (Allum et al., 1993, 1994) where different combinations of platform translation and tilt, designed to produce similar ankle dorsiflexion amplitudes, produced significantly different kinetic responses in the different body segments. Here, when discussing if the response to one stimulus is influenced by the presence of the other, only the FRF associated with such stimulus is considered, e.g. testing if the part of the body sway response explained by a linear transfer function applied to the support surface translation is affected by the presence of the tilt.

### 2.4. Statistical Analysis

To assess the effect of the superposition of support surface tilt and translation, the statistical significance of the observed difference between FRFs is tested using bootstrap hypothesis tests (Zoubir and Boashash, 1998). The norm of the difference between the averaged FRFs of the group of trials associated with the two conditions is compared versus the p<0.05 significance threshold for the null hypothesis that the two distributions have the same average. Hence, in the following, a condition is considered to produce a significant effect if that effect is outside of the corresponding 95% confidence interval. Each comparison between two groups of FRFs was tested separately. A thorough description of the performed tests is reported in the Supplementary Material.

### 2.5. Modeling and simulation

The simulation aims to test qualitatively the hypothesis that the observed differences can be explained by a different joint stiffness occurring with the different combinations of tilt and translation. The underlying rationale is that the rejection of support surface tilt benefits from joint compliance, while the opposite applies to support surface translation since this is compensated by joint stiffness. In previous works (Allum et al., 1993) no difference in ankle stiffness was suggested by the analysis of joint torque and muscle activation when the tilt with small amplitude was superimposed on translation. Also here we propose the idea that stiffness is modulated according to the relative amplitude of the stimuli, as shown in Fig. 5 B.

The nonlinearity in the compensation of support surface tilt (Hettich et al. 2011, 2013, 2015) may partially explain the effect of the tilt on the FRF, as it produces oscillations with frequencies that are not present in the input and hence can overlap with the power-spectrum of support surface translation. To explain that there is an effect exerted by support translation on the tilt FRF, the passive torque from the biomechanical properties of the muscles and tendons (Flash & Hogan, 1985; Hogan, 1984) should additionally be taken into account. In particular, one may hypothesize an increase in ankle passive stiffness in response to the support surface translation, as corresponding co-contractions have been observed using electromyography (Henry et al., 1998). More refined modeling of muscle stiffness dynamics, as presented in De Groote et al. (2017), is beyond the scope of the present work, because this is focused on the analysis of the steady-state behavior.

A scheme of the simulation system used is shown in Fig. 5A, and its parameters are recapped in Fig. 5B. A single inverted pendulum (SIP) model similar to the one presented in Akçay et al. (2021) was used to simulate the scenario when comparing the different combinations of the two stimuli with corresponding EC conditions. The simulation was adapted to include here, however, support surface tilt, but with the control feedback being based on a body-in-space sway signal affected by the aforementioned dead-zone nonlinearity for the velocity of foot-in-space rotation (Fig. 5C). The non-linearity represented by trigonometric functions, as they are involved in the dynamics of the system (e.g. the sine of the body sway defining gravity torque), is here negligible because of the small angles considered for upright stance (e.g. BS << 10°). In general, upright balance is a multiple degrees of freedom task (Alexandrov et al., 2017, 2015; Duchene et al., 2021; Glass et al., 2022). However, the SIP model, although not considering hip-ankle coordination explicitly, can provide a good prediction of the control of the body's center of mass (COM) both in quiet stance (Morasso et al., 2019; Morasso and Schieppati, 1999; Winter et al., 1998) and perturbed conditions (Hettich et al., 2011; Peterka, 2002). This means that the SIP model can here be used to simulate the sway of the body COM with respect to the ankle joint, although this is involved in ankle-hip coordination. Notably, the posture control model proposed in (Hettich et al., 2011) is a double inverted pendulum, but the ankle joint is controlled with a feedback system involving the sway of the COM. In effect, this mimics controlling a virtual SIP that produces similar results in terms of body sway, as it is obtained when using directly a SIP model.

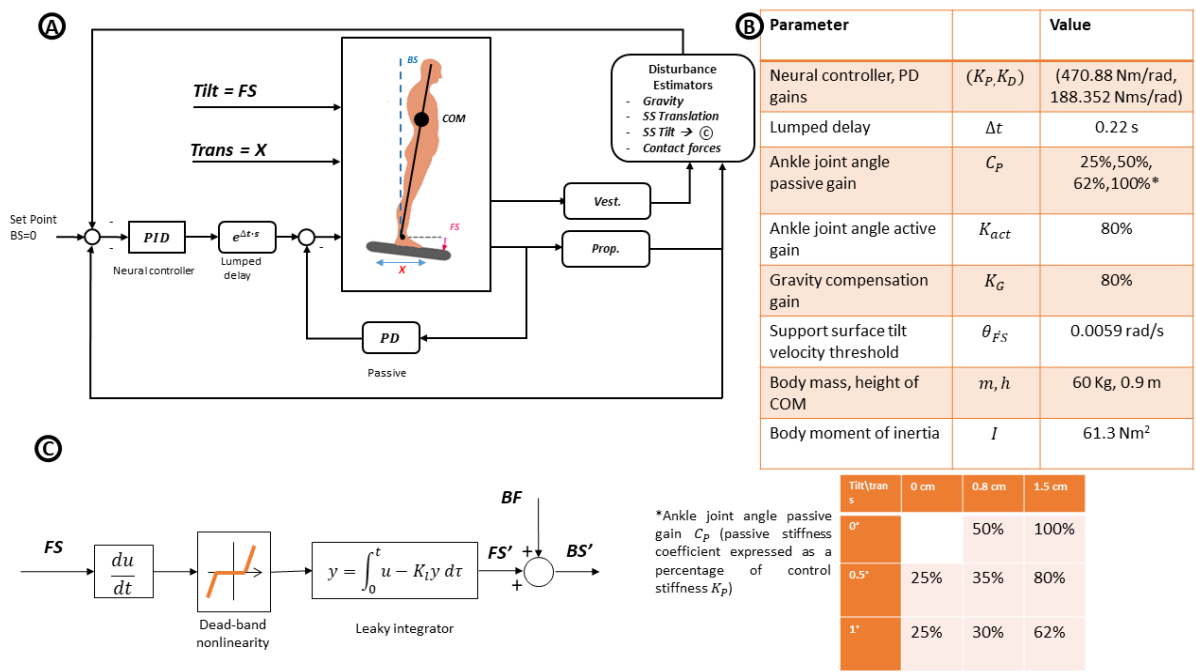

*Fig. 5.* *The model used for the simulations. A. Overall control scheme. The disturbance estimators include the support surface tilt estimator shown in C. B. Simulation parameters. The passive part, i.e. the ankle joint passive stiffness, defined by the gain $C_P$, changes according to the simulated scenario, as shown in the table below. C. Nonlinearity affecting the estimation of body-sway-in-space (BS') within the support surface (SS) tilt estimator. The nonlinearity consists of a deadband applied on the SS tilt velocity estimate of foot-in-space (FS). Further, the estimation BS' is reconstructed involving the body-to-foot angle (BF) provided by the ankle proprioception.*

## 3. Results
### 3.1. The direct effect of vision

A comparison between the results obtained for the EO and EC conditions showed significant differences (p>0.05) for all tested tilt and translation conditions (for a total of 12 comparisons). More in detail, the difference between EC versus EO responses was evaluated for all combinations of the two stimuli. Considering the response to support surface tilt, the averaged FRFs in Fig. 3 show that the EC condition is associated at higher frequencies with relatively larger gain values. And, support surface translation with EC produces a response that at the low frequencies has a higher gain than the corresponding EO response (see Fig. 4). The significant effect of vision availability confirms the results of previous studies e.g. (Akçay et al., 2021; Assländer et al., 2015; Jilk et al., 2014)

## 3.2. Superposition of disturbances

Testing for the effects of different stimulus combinations, comparisons were performed across two different conditions of the additional stimulus (the one that is not considered as input in the FRF) while leaving all the other conditions unchanged, this lead to a total of 24 comparisons.

There was no statistically significant effect of superimposing the stimuli ($p<0.05$) except in 5 of all the cases tested. Specifically, these were:

1. EC tilt of 0.5° without translation *vs* EC tilt of 0.5° with a 1.5 cm translation (p=0.0002)
2. EC tilt of 1° without translation *vs* EC tilt of 1° with a 1.5 cm translation (p=0.0166)
3. EO tilt of 0.5° without translation *vs* EO tilt of 0.5° with a 1.5 cm translation (p=0.0031)
4. EO translation of 0.8 cm without tilt *vs* EO translation of 0.8 cm with a tilt of 1° (p=0.0467)
5. EC 0.8 cm translation without tilt *vs* EC 0.8 cm translation with a tilt of 1° (p=0.0378)

With EC, a significant superposition effect is observed also with the 1° stimuli (case 2). This may be because the effect of translation with EC tends to be overall larger.

## 3.3. Effect on leg and trunk sway responses

In §2.2 it was shown how the COM sway is a good descriptor of the postural behavior in the presented scenario because it is strongly correlated with ankle angle. However, the two angles are not exactly the same (see Fig. 2, bottom). In order to test the effect of the stimuli on the two body segments specifically, the test presented in the previous section has been applied to the FRFs of legs sway and trunk sway.

The effect of stimuli superposition was significant in the following cases:

1. EO tilt of 0.5° without translation vs EC tilt of 0.5° with a 1.5 cm translation (p=0.00795)
2. EC tilt of 0.5° without translation vs EC tilt of 0.5° with a 1.5 cm translation (p=0.000, i.e. out of the histogram)
3. EO tilt of 1° without translation vs EC tilt of 1° with a 1.5 cm translation (p=0.0109) *
4. EC 0.8 cm translation without tilt vs EC 0.8 cm translation with a tilt of 1°. (P=0.01755)

Case 3 was not producing significant effects on COM sway, but it affected trunk sway (see below). There were two cases associated with significant effects on COM sway but not on leg sway:

i. EC tilt of 1° without translation vs EC tilt of 1° with a 1.5 cm translation (p=0.0166, for LS it was 0.05145, not far from the threshold)
ii. EO translation of 0.8 cm without tilt vs EO translation of 0.8 cm with a tilt of 1° (p=0.0467, for leg sway it was 0.065, not far from, although beyond, the .05 threshold)

The effect of translation on tilt response was significant in the following cases all considering the effect of the translation stimulus on the user, in all cases the effect of translation was observed on the response to tilt:

1. EC tilt of 1° without translation vs EC tilt of 1° with a 1.5 cm translation (p=0.0199)
2. EC tilt of 0.5° without translation vs EC tilt of 0.5° with a 1.5 cm translation (p=0.000, i.e. out of the histogram)
3. EO tilt of 0.5° without translation vs EC tilt of 0.5° with a 1.5 cm translation (p=0.000)
4. EO tilt of 0.5° without translation vs EC tilt of 0.5° with a 0.8 cm translation (p=0.0177)
5. EO tilt of 0.5° ° with a 0.8 cm translation vs EC tilt of 0.5° with a 1.5 cm translation (p=0.0138)
6. EO tilt of 1° without translation vs EC tilt of 1° with a 1.5 cm translation (p=0.0104)

There was no significant effect of the tilt stimulus on the trunk response to translation. This is coherent with the fact that SS translation has a larger effect on TS than on LS (Lippi et al., 2020), while the opposite stands for SS tilt (Hettich et al., 2014, Fig, 3).

The results of the tests are shown in detail in the additional materials.

## 4. Simulation Results

The results of the simulations, which were obtained using the model in Fig. 5A, are shown in Fig. 6. A change in the ankle joint stiffness induced by the translation is assumed to produce also some change in the response to tilt (see Fig. 6, top row). The effect of the nonlinearity applied to the tilt stimulus is reflected in the FRF relative to translation. This effect is larger, as the tilt amplitude grows larger compared to the translation amplitude (Fig. 6). The differences between the FRFs relative to 0.5° and 1° tilt stimulus (Fig. 6, top, left vs right) are due to the nonlinearity affecting support surface tilt sensing (Fig. 4. C).

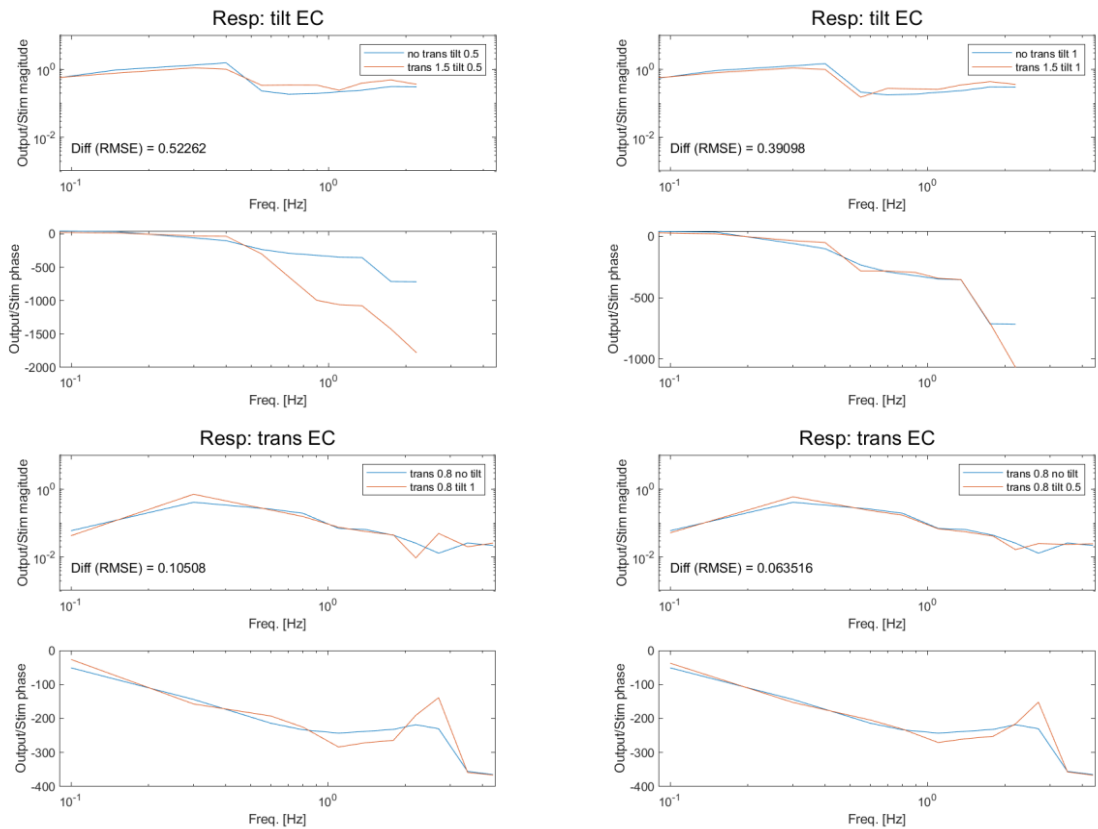

*Fig. 6. Frequency response functions (FRFs) obtained with simulations of the model shown in Fig. 6A. In each plot, two different conditions are compared. Specifically, it is shown how the responses to the smaller tilt and translation stimuli (i.e. 0.5° and 0.8 cm respectively) are affected by the presence of the other disturbance modality. There is an observable effect on the tilt FRF due to the different ankle stiffness induced by the translation (and imposed in the simulation by the choice of parameters shown in the table in Fig. 4). The translation FRF is affected by oscillations induced by the tilt stimulus outside of the domain of the input power-spectrum due to the non-linearity: the effect is larger when the tilt is larger compared to the translation (compare with Fig. 13). The term "Diff (RMSE)" quantifies the difference between the two plotted FRF in the complex domain as the root mean square of the difference between the 11 components.*

## 5. Discussion

The results show a difference between the EC responses to the support surface rotations without versus with the additional presence of a support surface translation of 1.5° amplitude. Similarly, with EO a pronounced difference is observed for support surface tilt with the amplitude of 0.5°. In contrast, translation with the 1.5 cm amplitude was hardly affected at all by the presence of support surface tilt (at least, a difference of relevant magnitude was not observable). A considerable difference, however, was observed between the cases without tilt versus the case with the largest tilt, i.e. 1°. This condition is similar to the one that produced significant differences in the tilt responses where the FRF is defined as the ratio between two power-spectra (§2.2). Noticeably, with the power-spectrum of the input as the denominator, the effect of external disturbances increases when the input becomes smaller.

A specific point of the present work consists of the very small amplitude of the disturbances. Interestingly, the smallest stimulus amplitude (0.5° tilt) was the one that showed a difference in the response when induced by the presence of the other disturbance modality (1.5 cm translation). This suggests that, with larger amplitudes, the responses to the two stimuli would become independent, at least up to those amplitudes where we observed the onset of a strong nonlinearity (e.g. when the total sway induced by the two disturbances forced the subject to perform a recovery step). Our future work will explore the effect of such large stimuli.

Those cases where the superposition of the extra stimuli did not induce a considerable effect on the response to the stimulus under investigation suggest that tilt and translation can be tested simultaneously given the right choice of stimulus amplitudes. Currently, the here presented data set is used to define benchmark tests for the evaluation of the posture control of humanoids (Lippi et al., 2019, 2021; Torricelli et al., 2020). Both the trials with and without superposition were used: this increased the number of samples available for the statistics.

We expect that testing responses to support surface tilt and translation not only separately, but also simultaneously provides advantages for system identification as observed in previous studies on technical systems (Peres et al., 2014) and biological systems (Westwick et al., 2006). As to the specific cases tested here, including support surface tilt in the tests of translation stimuli makes the relative contribution of passive and active forces less ambiguous compared to the use of support surface alone, since the two evoked torque effects have a similar profile when BF and BS angles are the same (i.e. the SS is not tilted).

The simulations suggested that the known nonlinearities affecting the response to support surface tilt do not produce a major interaction between the responses to the two stimulus modalities. Conversely, a change in ankle stiffness depending on the disturbance combination can produce a significant interaction between the responses to the two stimuli.

The fact that in most of the cases there was not a significant difference between the response to a stimulus when it is provided alone or in combination with an additional one is compatible with the idea that external disturbances are identified and responded to specifically, as suggested in previous work that assumed disturbance specific estimations and compensations (T. Mergner, 2010).

It is worth noticing that here the body sway steady-state response to continuous stimuli is expressed in terms of FRF. This description does not take into account possible differences in muscle activations (studied, for example in Allum et al., (1993) and purposely aims to cut out transient effects. This means that there could be differences induced by the different combinations of stimuli that are not reflected in the examined behavior. Furthermore, the behavior induced by continuous stimulation is characterized by specific dynamics that may differ from what is observed in fast transient responses. For example, a model simulation with the same parameters used in reproducing the steady-state FRFs with relative fast foot dorsiflexion and support translation produces the responses shown in Fig. 7, which look relatively slow (e.g. compare with Allum & Honegger, 1992, Fig 1).

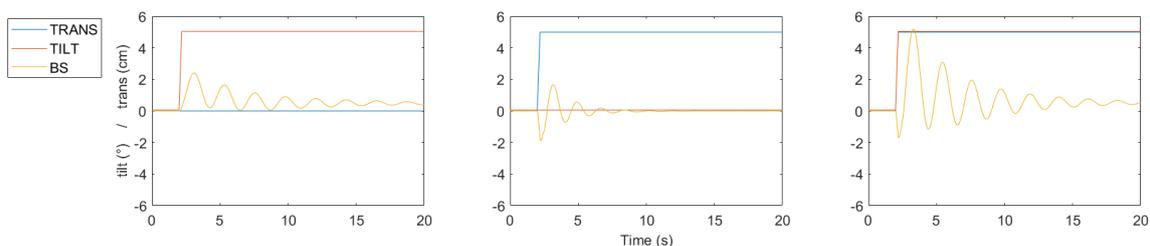

**Fig. 7** *Simulation model, responses to fast ramps (5° and 5 cm respectively, in 100 ms). Small oscillations are produced for several seconds. The passive gain is set to 100% for translation alone and 35% for the superposition, replicating the scenarios described in Fig. 5.*

A limitation of the current study is that both the perturbation stimuli act in the sagittal plane. Previous works have shown that stimuli presented in different planes within the same modality (e.g. pitch and roll rotations) applied with different delays, can produce interactions reflected by significantly different responses among the different combinations of stimuli, see, for example, Grüneberg et al. (2005) and Küng et al. (2009). Specifically, in there is an asymmetry in the response in that the response to support surface pitch is influenced by the presence of the roll stimulus while the opposite does not happen significantly. The DEC control used in this work was implemented independently in the control of the frontal and sagittal plane of a humanoid robot (Lippi & Mergner, 2017). Interestingly the responses to PRTS support surface tilt in a 45° plane were larger than the ones in the sagittal plane given the same amplitude, this can be explained by an effect of the gain nonlinearity (see Fig. 5C) that makes the components of the stimulus in the frontal and the sagittal plane be less compensated (because they are smaller) than a stimulus of the same amplitude applied just on one of the planes. Human experiments in this scenario are to be performed. The present work was focused on COM sway, but interestingly, as shown in §3.3, the superposition of the stimuli affects differently the trunk sway response to translation and tilt. This suggests future work to explore such effects with a model accounting for more degrees of freedom.